\documentclass[aps,11pt,showpacs,onecolumn,nofootinbib]{revtex4}

\usepackage[utf8]{inputenc}
\usepackage{array}
\usepackage[english]{babel}
\usepackage[T1]{fontenc}
\usepackage{amsmath}
\usepackage{amsfonts}
\usepackage{amssymb}
\usepackage{kpfonts}
\usepackage{hyperref}
\usepackage{subeqnarray}
\usepackage{tensor}
\usepackage{graphicx}
\usepackage{color}

\DeclareMathOperator{\E}{e}

\begin{document}

\title{Straight spinning cosmic strings in Brans-Dicke gravity}

\author{{S. Mittmann dos Santos}}
\email{sergio.santos@poa.ifrs.edu.br}
\affiliation{Universidade Estadual Paulista Júlio de Mesquita Filho -- UNESP\\
Câmpus de Guaratinguetá, 12.516-410, Guaratinguetá, SP, Brazil}
\affiliation{Instituto Federal de Educação, Ciência e Tecnologia do Rio Grande do Sul -- IFRS\\
Câmpus Porto Alegre, 90.030-041, Porto Alegre, RS, Brazil}

\author{{J. M. Hoff da Silva}}
\email{hoff@feg.unesp.br}
\author{{J. L. Cindra}}
\email{lourenco@feg.unesp.br}
\affiliation{Universidade Estadual Paulista Júlio de Mesquita Filho -- UNESP\\
Câmpus de Guaratinguetá, 12.516-410, Guaratinguetá, SP, Brazil}

\begin{abstract}
It is presented an exact solution of straight spinning cosmic strings in Brans-Dicke theory of gravitation. The possibility of the existence of closed timelike curves around these cosmic strings is analyzed. Furthermore, the stability about the formation of the topological defect discussed here is checked. It is shown the existence of a suitable choice for the integration constants in which closed timelike curves are not allowed.
We also verify if it is possible that the obtained spacetime can be the source that describes the observed rotational curves in some galaxies.
\end{abstract}


\maketitle

\section{Introduction}

The Brans-Dicke (BD) gravitation is the simplest scalar-tensorial theory in which the gravitational phenomena are described by a tensorial and a scalar field \cite{Brans1961}. The dynamical equations, for the scalar and tensorial fields, have a dimensionless parameter, $\omega$, controlling the departure from usual General Relativity (GR). The closer the unit is this parameter, the greater is the difference between the description of phenomena given by GR and the BD theory. Current solar system experiments sets $|\omega|>40000$ \cite{limitexperimental}. Nevertheless, scalar-tensorial theories of gravitation appears somewhat naturally in the development of unified theories of fundamental interactions, since grand unification models require of a spin-0 field in order to explain the gravitational phenomena \cite{Green1987}. Moreover, a seminal result coming from the investigation of the cosmological evolution of the Universe, under the auspices of BD theory, shows GR as a strong attractor of BD gravitation in the sense that the dynamics of the scalar field has been gradually suppressed \cite{Damour1993, Damour1993A, Contaldi1999}.

It is usually understood that cosmic strings are topological defects, which arose from a spontaneous symmetry breaking occurred in some phase transition of the early Universe, and therefore can provide some information about most fundamental theories. Spinning cosmic strings are a particular class of approximately one-dimensional topological defects that have an angular velocity around the longitudinal axis of symmetry and whose dynamics is associated to the G\"odel's solutions \cite{Godel1949, Novello1983}. Spinning strings were largely studied in GR, and its solutions confirm the existence of closed timelike curves (CTC's) in at least part of the resulting spacetime around the defect \cite{Jensen1992}. The point to be emphasized in those previous works is that the CTC's are generally generated by non-physical sources \cite{Jackiw1992}. As the formation of spinning cosmic strings would have occurred in the early Universe, an approach with BD gravitation, in the light of Refs. \cite{Damour1993, Damour1993A}, can bring some additional information not yet obtained within GR. In Ref. \cite{Anchordoqui1999} spinning cosmic strings were investigated in the context of BD gravity, but the obtained solutions are overly particular, since the performed analysis takes advantage of a scalar field particular shape.

In this paper we shall delve into the question of the spacetime generated by a straight spinning cosmic string within BD gravity, outside the spinning string. We show the complete solution for the metric plus scalar field system and study the physical prospects. It is important to remark that by studying the exterior spacetime solution, the connection with BD gravity is weakened and most part of the solution is obtained in the framework of a scalar field minimally coupled to gravity. We shall still maintain the presentation within the BD scope, preparing the solution to a further approach taking into account the interior solution as well. However, in order to keep track with BD theory we make an effort mentioning the behavior in the Jordan frame, and evincing possible values of the integration constants particular to the scalar sector. In this vein, we use the so-called Einstein frame to perform all the calculations, but after all it is shown that the general qualitative behavior can also be applied to the Jordan frame as well, exception made to the general shape of the scalar field. The physical outputs to the emphasized here are the regularity of the spacetime outside the spinning cosmic string evinced by the Kretschmann scalars, the existence of a radial proper distance, say $r^*$, from the string below which CTC's are (in principle) possible, and the analysis of energy quantization of a non inertial particle surrounding the spinning string pointing to the physical instability of the source. Moreover it is presented a particular choice of the integration constants in which CTC's are not allowed. This last fact coming exclusively from the scalar coupling.
Finally, it is verified the possibility that a spacetime like the one obtained here can describe the rotational curves of the analyzed galaxies in Ref. \cite{Rubin1980}, as an alternative to dark matter.
In the rest of this introductory section we shall fix the notation.

The gravitational action in BD theory is given by
\begin{equation}
S=\frac{1}{16 \pi} \int d^4 x \sqrt{-\bar{g}} \left ( \Phi \bar{R} -\frac{\omega}{\Phi} \partial^\mu \Phi \partial_\mu \Phi \right )+ S_\textit{matter} \; ,
\label{acao_Brans-Dicke}
\end{equation}
where $\omega$ is the aforementioned dimensionless parameter. The equations of motion in the Jordan frame read
\begin{eqnarray}
G_{\mu\nu}&=&\frac{\omega}{\Phi^2} \left ( \partial_\mu \Phi \partial_\nu \Phi - \frac 1 2 \bar{g}_{\mu\nu} \bar{g}^{\rho \sigma} \partial_{\rho} \Phi \partial_\sigma \Phi \right ) +\frac 1 \Phi \left ( \partial_\mu \partial_\nu \Phi - \bar{g}_{\mu\nu} \bar{\Box} \Phi \right )+ \frac{8\pi}{\Phi}T_{\mu\nu} \; , \nonumber \\
\bar{\Box} \Phi &=& \frac 1 {\sqrt{-\bar{g}}} \partial_\mu \left ( \sqrt{-\bar{g}} \partial^\mu \Phi \right )= \frac {8\pi}{2\omega +3} T \; .
\label{eqs_movimento_BD_Jordan}
\end{eqnarray}
Notice that we have reserved the notation $\Phi$ and bar terms for Jordan frame evaluated quantities. There is a more convenient way to work with the BD equations, which is through the Einstein's frame. In this frame, $\bar{g}_{\mu\nu}$ and $\Phi$ from the Eqs. (\ref{eqs_movimento_BD_Jordan}) are redefined by the new dynamical variables $g_{\mu\nu}$ and $\phi$, as in \cite{Boisseau1998}:
\begin{eqnarray}
\bar{g}_{\mu\nu}& = & \E^{ 2\kappa \phi } g_{\mu\nu} \; ,\nonumber \\
\Phi& = & \frac 1 G \E^{ -2 \kappa \phi } \; ,
\label{variaveis_Jordan-Einstein}
\end{eqnarray}
where
\begin{equation}
\kappa^2=\frac 1 {2 \omega + 3}
\label{kappa}
\end{equation} and $G$ is a factor related to the Newton gravitational constant. Thus, in the Einstein's frame, the Eqs. (\ref{eqs_movimento_BD_Jordan}) can be written as
\begin{eqnarray}
G_{\mu\nu}&=& 2 \partial_\mu \phi \partial_\nu \phi - g_{\mu\nu} g^{\rho \sigma} \partial_{\rho} \phi \partial_\sigma \phi + \frac{8\pi}{\phi}T_{\mu\nu} \; , \label{eqs_movimento_BD_Einstein_1}
 \\
\Box \phi &=&
\frac {8\pi}{2\omega +3} T \; .
\label{eqs_movimento_BD_Einstein_2}
\end{eqnarray} As we can see, the appreciation of BD field equations in the Einstein frame leads to a type of decoupling of the scalar and tensorial modes in the vacuum. There are a large amount of works dealing with the subtle aspects of physics considered from the point of view of these two frames (for a broader review see \cite{Faraoni1999}, and for a relatively new advance in the issue see \cite{Quiros2013}). For the system we are about to consider, since the qualitative behavior is basically the same in both frames, we shall perform and interpret the results in the manageable Einstein frame and only occasionally make reference to the Jordan frame.

This paper is organized as follows: in the next section we obtain the solution for a straight spinning cosmic string and the scalar field in the Einstein frame of BD gravity, outside the string, in a quite general form. In Section III we study the physical content of the solution. In the final section we conclude.

\section{Solution for a spinning cosmic strings in Brans-Dicke theory}

The spinning cosmic string to be addressed in this work is considered as very small radius $R$ ($R \rightarrow 0$) source along with a very large length $L$ ($L \rightarrow \infty$). The more general cylindrically symmetric line element outside such a defect reads \cite{Jensen1992}
%
\begin{equation}
ds^2 = -\left( \E^\alpha dt+M \; d\varphi \right)^2+r^2 \E^{-2\alpha} d\varphi^2 + \E^{2 \left( \beta-\alpha \right) } \left( dr^2+dz^2 \right) \; ,
\label{metrica}
\end{equation}
where the angular coordinate varies, in principle, in the range $0\leq\varphi \leq 2\pi$. The string is then placed along the $z$-axis at $r=0$ and, by means of the cylindrical symmetry we have $\alpha$, $\beta$ and $M$ as functions of the, strictly positive, radial distance $r$ only. We shall present the field equations by using the Cartan forms \cite{Soleng1990,Plebanski2006}, in which
\begin{equation}
ds^2=\eta_{ij} \theta^i \theta^j \; ,
\label{metrica_Cartan}
\end{equation}
where $\eta_{ij}$ is the Minkowski metric and $\theta^i$ is the tetrad basis
\begin{equation}
\theta^i=\theta^i_{\;\;\mu}dx^\mu \; ,
\label{tetrad_basis_2}
\end{equation}
with
\begin{equation}
\theta^t=\E^\alpha dt+M d\varphi \; , \;\;\;\;\;\;\;\;\;\;
\theta^ r=\E^{\beta-\alpha} dr \; , \;\;\;\;\;\;\;\;\;\;
\theta^\varphi=r \E^{-\alpha} d\varphi \; , \;\;\;\;\;\;\;\;\;\;
\theta^z=\E^{\beta-\alpha} dz \; ,
\end{equation}
\begin{equation}
\theta^t_{\;\;t}=\E^\alpha \; , \;\;\;\;\;\;\;\;\;\;
\theta^t_{\;\;\varphi}=M \; , \;\;\;\;\;\;\;\;\;\;
\theta^r_{\;\;r}=\E^{\beta-\alpha} \; , \;\;\;\;\;\;\;\;\;\;
\theta^\varphi_{\;\;\varphi}=r \E^{-\alpha} \; , \;\;\;\;\;\;\;\;\;\;
\theta^z_{\;\;z}=\E^{\beta-\alpha}
\label{transformacao_Cartan}
\end{equation}
and
\begin{equation}
dx^\mu=\theta_i^{\;\;\mu} \theta^i \; ,
\end{equation}
where
\begin{equation}
\theta_i^{\;\;\mu}=\eta_{ij}\; g^{\mu\nu}\theta^j_{\;\;\nu} \; .
\label{eq_9-12_Plebanski}
\end{equation}
When torsion is absent, the Cartan's first structure equation is
\begin{equation}
d\theta^i=-\omega^i_{\;\;j}\wedge\theta^j \; .
\label{1a_eq_Cartan}
\end{equation}
By means of the metricity condition and because $\omega_{ij}$ is antisymmetric, the non zero components of the connection forms $\omega^i_{\;\;j}$ are \cite{Soleng1990}
\begin{eqnarray}
\omega^t_{\;\;r} = \omega^r_{\;\;t} &=& \alpha'\E^{\alpha-\beta}\theta^t-\Omega\theta^\varphi \; , \nonumber \\
\omega^t_{\;\;\varphi} = \omega^\varphi_{\;\;t} &=& \Omega\theta^r \; , \nonumber \\
\omega^r_{\;\;\varphi} = -\omega^\varphi_{\;\;r} &=& -\Omega\theta^t + \left( \alpha'-\frac 1 r \right) \E^{\alpha-\beta} \theta^\varphi \; , \nonumber \\
\omega^r_{\;\;z} = -\omega^z_{\;\;r} &=& \left( \alpha'-\beta' \right) \E^{\alpha-\beta} \theta^z \; .
\label{connection_forms}
\end{eqnarray}
In the above set of equations, a prime means derivative with respect to $r$ and $\Omega$ denotes
\begin{equation}
\Omega= \frac{M\alpha'-M'}{2r} \; \E^{2\alpha-\beta} \; .
\label{Omega}
\end{equation}
The curvature forms $\Omega^i_{\;\;j}$ are determined by the second Cartan's structure equation
\begin{equation}
\Omega^i_{\;\;j}=d\omega^i_{\;\;j}+\omega^i_{\;\;k}\wedge\omega^k_{\;\;j} \; ,
\label{2a_eq_Cartan}
\end{equation}
which are related to the Riemann tensor $\mathcal{R}^i_{\;\;jkl}$ by
\begin{equation}
\Omega^i_{\;\;j}=\frac 1 2 \mathcal{R}^i_{\;\;jkl} \; \theta^k \wedge\theta^l \; .
\label{tensor_Riemann-Cartan}
\end{equation} Obviously, when $i=k$, the contraction $\mathcal{R}^i_{\;\;jil}$ leads to the Ricci tensor $\mathcal{R}_{jl}$. Thus, within this context, the Einstein tensor $\mathcal{G}_{ij}$ reads
\begin{equation}
\mathcal{G}_{ij}=\mathcal{R}_{ij}-\frac 1 2 \eta_{ij}R \; ,
\label{Einstein_tensor_Cartan_form}
\end{equation}
where $R$ is the Ricci scalar $R=\eta^{ij}\mathcal{R}_{ij}$. The relation between (\ref{eqs_movimento_BD_Einstein_1}) and (\ref{Einstein_tensor_Cartan_form}) is given by \cite{Plebanski2006}
\begin{equation}
\mathcal{G}_{ij}=\theta_i^{\;\;\mu} \theta_j^{\;\;\nu} G_{\mu\nu} \; .
\label{Einstein_tensor_Cartan_form_2}
\end{equation}
Thus, outside the string, the dynamical equations (\ref{eqs_movimento_BD_Einstein_1}) and (\ref{eqs_movimento_BD_Einstein_2}) in the vacuum may be recast into the following form
\begin{eqnarray}
\mathcal{G}^t_{\;\;t}&:&-3\Omega^2-\left( 2\alpha''-\beta''+\frac{2\alpha'}{r}-\alpha'^2 \right) \E^{2 \left( \alpha-\beta \right)}
=-\E^{2 \left( \alpha-\beta \right)} \phi'^2
\label{sistema_eq_1} \; , \\
\mathcal{G}^r_{\;\;r}&:&\Omega^2+\left( \frac {\beta'}{r}-\alpha'^2 \right) \E^{2 \left( \alpha-\beta \right)}
=  \E^{2 \left( \alpha-\beta \right)} \phi'^2
\label{sistema_eq_2} \; , \\
\mathcal{G}^z_{\;\;z}&:&-\Omega^2+\left( \alpha'^2-\frac{\beta'}{r} \right) \E^{2 \left( \alpha-\beta \right)}
= -\E^{2 \left( \alpha-\beta \right)} \phi'^2
\label{sistema_eq_4} \; , \\
\mathcal{G}^\varphi_{\;\;\varphi}&:&\Omega^2+\left( \beta''+\alpha'^2 \right) \E^{2 \left( \alpha-\beta \right)}
= -\E^{2 \left( \alpha-\beta \right)} \phi'^2
\label{sistema_eq_3} \; , \\
\mathcal{G}^t_{\;\;\varphi}&:&\left [ \Omega'+2\alpha' \Omega+\Omega \left( \beta'-\alpha' \right) \right ] \E^{\alpha-\beta}
= 0
\label{sistema_eq_5} \; , \\
\Box \phi &:& \frac {\E^{2 \left( \alpha-\beta \right)} \phi'} 2 \frac {d}{dr} \left\{ \ln \left[\E^{4\left( \beta-\alpha \right)} r^2 \left( \E^{2 \left( \alpha-\beta \right)} \phi' \right)^2\right] \right\}
= 0 \; ,
\label{sistema_eq_6}
\end{eqnarray}
where the scalar field $\phi$ is also a function of $r$ (i.e., $\phi = \phi(r)$) only.

If the scalar field is non-dynamical, there exists a particular solution given by $\Omega=0$ and $\alpha'=0$
($M'=0$). In this case, it is possible to associate $M$ to (minus)
$J$ and $\E^{-\alpha}\sim 1-\mu/\left(2\pi\right)$, where $\mu$ and $J$ are the
mass and the spin of a spinning point mass \cite{Jackiw1992}. The
metric (\ref{metrica}) represents then, for $dz=0$, the spacetime
exterior to such a particle in $(2+1)$-dimensional gravity or, by
considering it as a section of the full line element (\ref{metrica}),
$\mu$ and $J$ can be interpreted as densities of a spinning cosmic
string in usual GR. In BD framework, however, this usual approach gives non longer a solution of the
system of equations.

From now on, we shall implement Lorentz invariance along the $z$-axis. It is performed by setting
\begin{equation}
\beta(r)=2\alpha(r) \; ,
\label{beta_para_corda_reta}
\end{equation} in the line element (\ref{metrica}). With this constraint, it is fairly simple to see, from the sum of Eqs. (\ref{sistema_eq_1}) and (\ref{sistema_eq_2}), that $\Omega=0$, leading immediately to
\begin{equation}
M=a_1 \E^\alpha \; ,
\label{1a_solucao_para_M}
\end{equation} where $a_1$ is an integration constant. The Eqs. (\ref{sistema_eq_2}) and (\ref{sistema_eq_3}) can be rewritten as
\begin{eqnarray}
\frac {2 \alpha'}{r}-\alpha'^2 &=& \phi'^2  \label{sistema_solucao_alpha_eq_1} \; , \\
2 \alpha''+\alpha'^2 &=& -\phi'^2 \label{sistema_solucao_alpha_eq_2} \; ,
\end{eqnarray} from which we arrive at
\begin{equation}
\alpha''+\frac {\alpha'}{r}=0 \; ,
\label{subtracao_sistema_solucao_alpha}
\end{equation} whose solution is given by
\begin{equation}
\alpha(r)=a_2 \ln r + a_3 \; ,
\label{solucao_para_alpha}
\end{equation} where $a_2$ and $a_3$ are constants. Now, substituting the solution (\ref{solucao_para_alpha}) back into Eq. (\ref{sistema_solucao_alpha_eq_1}), for instance, we have
\begin{equation}
\phi(r)=\phi_0 + \sqrt{a_2 \left( 2-a_2 \right) } \ln r \; ,
\label{solucao_para_phi}
\end{equation} where $\phi_0$ is a constant and we have discarded the negative branch of the solution. Besides, from (\ref{1a_solucao_para_M}), the $M$ function reads
\begin{equation}\label{solucao_para_M}
M(r)= a_1\E^{a_3}r^{a_2} \; .
\end{equation}
In order to have real scalar field the $a_2$ parameter must belong to the range
\begin{equation}\label{intervalo_a2}
0<a_2<2 \; .
\end{equation}
Note that at first sight, GR limit can be recover for $a_2=0$ or $a_2=2$, since the scalar field is constant for these values of $a_2$. Further analysis points to the fact that the limit to GR is better approached in the former case. Notice also, from Eq. (\ref{variaveis_Jordan-Einstein}) that in the Jordan frame the scalar field is well defined for every range of the non null radial coordinate\footnote{Moreover, had we choose the negative branch solution, the scalar field would be divergent as $r\rightarrow \infty$ in both, Einstein and Jordan, frames. In this context, a positive $\kappa$ along with the analysis in the Jordan frame seems to be more technically sound. Nevertheless some caution may be in order; after all, such an asymptotic behavior of the scalar field should not be taken so seriously since, in practice, at large distances other (different) sources should enter in the investigation and could even be more relevant.}. The obtained expressions for $\alpha(r)$, $M(r)$, and $\phi(r)$ are solutions of the system of Eqs. (\ref{sistema_eq_1})-(\ref{sistema_eq_6}) as far as $\beta=2\alpha$. Therefore the spacetime metric ends up as
\begin{eqnarray}
ds^2=-\E^{2a_3} r^{2a_2} \left( dt+a_1 d\varphi \right)^2+\E^{-2a_3} r^{2-2a_2} d\varphi^2 + \E^{2a_3} r^{2a_2} \left( dr^2+dz^2 \right) \; .
\label{metrica_corda_reta}
\end{eqnarray}

From the line element (\ref{metrica_corda_reta}), as well as the general shape of the scalar field, it is conceivable a note of warning about the spacetime behavior. The appreciation of the Kretschmann scalars, however, shows that the spacetime is well behaved in general. In fact, in the Einstein frame we have
\begin{eqnarray}
R^2=R_{\mu\nu}R^{\mu\nu}&=&4a_2^{\;2}\left(a_2-2\right)^2 \E^{-4a_3}r^{-4\left(1+a_2\right)}\; ,\nonumber\\
R_{\alpha\beta\mu\nu}R^{\alpha\beta\mu\nu}&=&4a_2^{\;2}\left(8-12a_2+7a_2^{\;2}\right) \E^{-4a_3}r^{-4\left(1+a_2\right)}\label{KEinstein}\; ,\\
C_{\alpha\beta\mu\nu}C^{\alpha\beta\mu\nu}&=&\frac{16}{3}a_2^{\;2}\left(1-2a_2\right)^2 \E^{-4a_3}r^{-4\left(1+a_2\right)}\; ,\nonumber
\end{eqnarray} where $C_{\alpha\beta\mu\nu}$ stands for the Weyl tensor. In view of the Eqs. (\ref{KEinstein}) the only subtle point is $r=0$, which is excluded from the analysis (recall that we are investigating the spacetime outside the string). It is indeed expected that the solution inside the string eliminates those (apparent) divergences. We reinforce that {\it mutatis mutandis} the general conclusions obtained here also apply to the Kretschmann scalars computed in the Jordan frame.

\section{Physical properties}

From the line element (\ref{metrica_corda_reta}) one sees that CTC's are in order, provided a negative $\varphi$ coefficient. For such it is straightforward to see that the inequality
\begin{equation}
\E^{-2 \alpha} \; r^2 - M^2 <  0
\label{condicao_CTC}
\end{equation}
must hold. This leads immediately to
\begin{equation}
r^{1-2a_2}< \E^{2a_3}a_1 \; . \label{NC}
\end{equation}
Therefore, we see that for $a_2=1/2$ it is possible to set a wide range of the parameters $a_1$ and $a_3$ such that CTC's are not allowed. This is a remarkable fact which have no counterpart in GR.

On the other hand, by assuming $a_1>0$ and $a_2 \neq 1/2$ the region allowing for the existence of CTC's is given by
\begin{equation}
r < r^*=\exp\left(\frac{2 a_3+\ln \; a_1}{1-2 a_2}\right) \; .
\label{r_qdo_ocorre_CTC}
\end{equation} We remark, by passing, that the same qualitative aspect is preserved in the Jordan frame. Eq. (\ref{r_qdo_ocorre_CTC}) delimits the cylindrical region in which, in principle, CTC's are possible. The radius of such a region depends whether $0<a_2<1/2$ or $1/2<a_2<2$, but qualitatively both situations are similar. This fact resembles the study of spinning strings in usual GR \cite{Jensen1992}. In the sequel we shall study a particular behavior of the geodesic motion, showing that in a purely radial motion, a test particle would move away from the CTC allowed region.

The geodesic equations derived from (\ref{metrica_corda_reta}) are given by
\begin{eqnarray}
r\ddot{t}+2a_1\left(2a_1-1\right)\dot{r}\dot{\varphi}+2a_2\dot{r}\dot{t}=0 \; , \nonumber\\
r\ddot{r}+a_2\left(\dot{r}^2+\dot{t}^2\right)+2a_1a_2\dot{t}\dot{\varphi}+\left[a_1^2a_2+\left(a_2-1\right) \E^{-4a_3}r^{2\left(1-2a_2\right)}\right]\dot{\varphi}^2
-a_2\dot{z}^2=0 \; ,\nonumber\\
r\ddot{z}+2a_2\dot{r}\dot{z}=0 \; ,\nonumber\\
r\ddot{\varphi}+2\left(1-a_2\right)\dot{r}\dot{\varphi}=0 \; ,\label{geodesics}
\end{eqnarray} where a dot stands for derivative with respect to $s$, the geodesic parameter. The complexity of Eqs. (\ref{geodesics}) is evident and its physical content is difficult to be highlighted. There is a particular case, nevertheless, which gives a clue on the radial behavior of the test particle. Hence, particularizing the motion in a $z$ and $\varphi$ constant plane, i. e., disregarding the angular and $z$-axis motion, the set of Eqs. (\ref{geodesics}) reduces to
\begin{eqnarray}
r\ddot{t}+2a_2\dot{r}\dot{t}=0 \; , \label{novageo1}\\
r\ddot{r}+a_2\dot{r}^2+a_2\dot{t}^2=0 \; . \label{novageo2}
\end{eqnarray} As far as $\dot{t}\neq0$ Eq. (\ref{novageo1}) may be written in the simple form
\begin{equation}
\frac{d\left[\ln\left(\,|\,\dot{t}\,|\,r^{2a_2}\,\right)\right]}{ds}=0 \; , \nonumber
\end{equation}
whose solution is given by $\dot{t}=C_1/r^{2a_2}$ being $C_1$ a positive constant. Therefore, substituting $\dot{t}$ into (\ref{novageo2}) we have
\begin{eqnarray}
r\ddot{r}+a_2\dot{r}^2+\frac{a_2C_1^{\,2}}{r^{4a_2}}=0 \; .\label{novageo3}
\end{eqnarray}
The radial motion of the test particle is, then, driven by Eq. (\ref{novageo3}). The solution of (\ref{novageo3}) is given in terms of the (inverse of) hypergeometric confluent functions. From the Figure 1, however, it is possible to extract its physical content. As we can see, the test particle moves away from the region with allowed CTC's.

\begin{figure}[h!]
\begin{center}
\includegraphics[width=12cm]{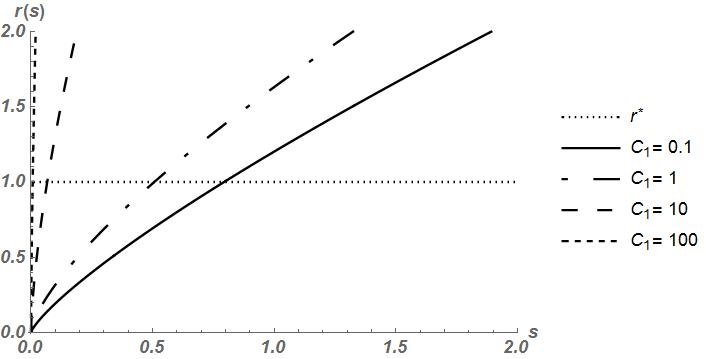}
\caption{Qualitative behavior of the radial coordinate. The constants are chosen such that $a_1=1$,  $a_2=\frac 1 4$ and $a_3=0$. For these values, the radius $r^*$ equals one (also shown in the figure).}
\label{fig_1}
\end{center}
\end{figure}

The previous analysis considers only the radial motion. The complement motion is also informative. Consider, for instance, a non inertial behavior at fixed non null radius, say $r=\bar{r}$. The quite naive coordinate transformation
\begin{eqnarray}
T &=& \E^{a_3} \bar{r}^{\;a_2} t+a_1 \E^{a_3}\bar{r}^{\;a_2} \varphi \; , \nonumber \\
Z &=& \E^{a_3}\bar{r}^{\;a_2} z \; , \label{transformacao_coordenadas} \\
\Phi &=& \E^{-a_3}\bar{r}^{\;1-a_2} \varphi \; , \nonumber
\end{eqnarray} turns the metric (\ref{metrica_corda_reta}) into a $(2+1)$-dimensional Minkowski type:
\begin{equation}
ds^2=-dT^2+dZ^2+d\Phi^2 \; .
\label{metrica_Minkowski}
\end{equation} Now, suppose a relativistic particle with mass $m$ subject to the gravitational field of the string but respecting the constraint $r=\bar{r}$. The Klein-Gordon equation for this particle shall present solutions of the form \cite{Mazur1986}
\begin{eqnarray}
\chi&=&\exp\left(-\,\frac {iET}{\hbar}\right) \; \exp\left(ikZ\right) \; \psi_{\bar{r}}(\varphi) \nonumber \\
&=& \exp\left(-\,\frac{iE \E^{a_3} r^{a_2}t}{\hbar}\right) \; \exp\left(-\,\frac{iEa_1 \E^{a_3}r^{a_2}\varphi}{\hbar}\right) \; \exp\left(ikZ\right) \; \psi_{\bar{r}}(\varphi) \; ,
\label{solucoes_chi}
\end{eqnarray} where $\psi_{\bar{r}} (\varphi)$ is a function determined by the constraint to be taking into account in the potential. Even without its explicit functional form, however, $\psi_{\bar{r}} (\varphi)$ must be a single-valued function of $\varphi$. For a complete turn around the string, the angular coordinate increases $2\pi$ and $T$ gives rise to a phase given by $\exp{\left(-2 \pi  i E a_1 \E^{a_3}\bar{r}^{\;a_2} /\hbar\right)}$. By adopting the interpretation that this lead to an energy quantization condition \cite{Mazur1987} (mathematically equivalent to periodic time boundary conditions) we are left with \begin{equation}
E = \frac{n \hbar}{a_1 \E^{a_3} \bar{r}^{\;a_2}}\equiv\frac{n \hbar}{M_{\bar{r}}} \; .
\label{quantizacao_energia_eq_2}
\end{equation}
It is therefore straightforward to see that for $\bar{r}$ far away enough from the string we shall have a very high value for $M_{\bar{r}}$. This qualitative analysis is particularly relevant since $M_{\bar{r}}/c^4$ is related to the periodicity of time (note that $M_{\bar{r}}/c^4$ may be expressed in seconds). Hence it indicates that the string is possibly unstable, as in GR. In this context, after its formation it was swept out due to instabilities. Again, the general aspects are also valid in the Jordan frame. Notice also that from (\ref{quantizacao_energia_eq_2}) one advocates that the right GR limit is given by $a_2=0$, since the standard energy quantization is recovered in this case \cite{Mazur1986}.


In the paper \cite{Santos2017}, a nearly cylindrically symmetric spacetime in the BD gravity was proposed,
as an alternative to dark matter, for the description of the rotational curves of the analyzed galaxies in Ref. \cite{Rubin1980}.
The model consisted of using the metric (\ref{metrica}) considering $\beta/\alpha\approx 2$, which leads to the violation of Lorentz invariance (unlike what was done here, when $\beta/\alpha = 2$, and the Lorentz invariance is preserved).
The stars of Ref. \cite{Rubin1980} were considered test particles moving in a circular trajectory around the centers of galaxies, in their equatorial planes.
The Lagrangian that describes this movement in the spacetime generated by the metric (\ref{metrica}) can be written as \cite{Matos2014}
\begin{equation}
2 \mathcal{L}=
-\;\E^{2 \alpha}\dot{t}^2 - M^2 \dot{\varphi}^2 -
 2 \E^\alpha M \dot{t} \dot{\varphi} +
 r^2 \E^{-2 \alpha} \dot{\varphi}^2 +
 \E^{2 \left( \beta - \alpha\right)} \dot{r}^2 +
 \E^{2 \left(\beta - \alpha\right)} \dot{z}^2 \; ,
\label{Lagrangiana}
\end{equation}
where dot stands for the derivative with respect to the proper time $\tau$.
From the Lagrangian (\ref{Lagrangiana}), it was determined that the ratio between the tangential velocity $v_\varphi$ of the stars and the velocity of light is
\begin{equation}
{ \left( \frac{v_{\varphi}}{c} \right) }_{\Omega_{\pm}M_{\pm}}=
\frac{\E^{-2\alpha}}{r}\,\left| \, \left( \E^{2\alpha}M_\pm^{\;2}-r^2 \right) \, \left( \Omega_\pm - \frac{\E^{3\alpha}M_\pm}{r^2\,-\,\E^{2\alpha}M_\pm^{\;2}} \right)\,\right| \; .
\label{velocidade_tangencial}
\end{equation}
In Eq. (\ref{velocidade_tangencial}), $\alpha$ and $M_\pm$ are the solutions for the functions of the metric (\ref{metrica}), obtained from the resolution of the system (\ref{sistema_eq_1})-(\ref{sistema_eq_6}), when $\beta/\alpha \neq 2$, and $\Omega_\pm$ is the angular velocity
\begin{equation}
\Omega_{\pm}=
-\;\frac{2\E^{2\alpha}\alpha'}{\E^\alpha \left( M'+M\alpha' \right)\pm \sqrt{\E^{2\alpha}\left( M'-M\alpha' \right)^2 \,-\,4r\alpha'\left( r\alpha' -1 \right)}} \; .
\label{velocidade_angular}
\end{equation}
As $r$ grows, the tangential velocities tend to the constant value
\begin{equation}\label{vvarphicstc}
{v_\varphi}_{cst}= c \,\sqrt{\frac{1-k_1}{1+k_1}} \; ,
\end{equation}
where
$k_1$ is one of the constants of the functions $\alpha$ and $M$.
The Eqs. (\ref{velocidade_tangencial}) and (\ref{vvarphicstc}), with adequate values for the integration constants of the solutions for the functions of the system (\ref{sistema_eq_1})-(\ref{sistema_eq_6}),
reflect the behavior of the reported measures in Ref. \cite{Rubin1980}.

When the symmetry is cylindrical, as adopted here, and not nearly cylindrical, as in Ref. \cite{Santos2017}, the application of the solutions (\ref{beta_para_corda_reta}), (\ref{solucao_para_alpha}), (\ref{solucao_para_phi}) and (\ref{solucao_para_M})
in Lagrangian (\ref{Lagrangiana}) shows that the ratio between the tangential velocity and the velocity of light is
\begin{equation}\label{vvarphic_corda_reta}
\left( \frac{v_\varphi}{c}\right)_\pm= \left | \, \frac{a_2 r\,\pm\, a_1\E^{2a_3} r^{2a_2}\sqrt{a_2\left( 1-a_2\right)}} {r\sqrt{a_2\left( 1-a_2\right)}\,\pm\, a_1 a_2 \E^{2a_3} r^{2a_2}}\, \right | \; ,
\end{equation}
which does not tend to a constant value, as expected for the description of the rotational curves.
There is an alternative that makes the ratio (\ref{vvarphic_corda_reta}) constant, what is when $a_2= 1/2$ (exactly the value of the constant $a_2$ so that CTC's do not occur, according to the condition (\ref{NC})),
but this leads to ${v_\varphi}_{cst}=c$, because the numerator of the Eq. (\ref{vvarphic_corda_reta}) is equal to its denominator for any $r$ (including for small $r$, which also does not agree with what is observed).
Therefore, a spacetime as defined by Eqs. (\ref{beta_para_corda_reta}), (\ref{solucao_para_alpha}), (\ref{solucao_para_phi}) and (\ref{solucao_para_M}) can not be a source of the rotational curves of the galaxies studied in Ref. \cite{Rubin1980}.

\section{Final Remarks}

It was performed a complete analysis of the outer spacetime generated by spinning strings within the context of the simplest scalar-tensor theory. The obtained solution is not based on a particular form of the scalar field and is completely characterized by means of the Kretschmann scalars it generates.

As we have seen, straight spinning cosmic strings are sources of CTC's in BD gravity, bringing some resemblance with solutions in standard GR. Moreover we have shown that the strictly radial geodesic motion is such that the test particle is driven away from the CTC's allowed region.

The very existence of CTC's is, obviously, an intriguing aspect related to these sources. In fact, the cylindrical region comprised by $r<r^*$ is a causality-violating one. In this vein, the analysis leading to the energy quantization is relevant. In a similar fashion of what was shown in GR \cite{Mazur1986}, we also concluded that, with the modeling of a infinitely thin spinning string performed by the BD theory of gravitation, the existence of such sources seems improbable.

It is important to elaborate on the peculiar possibility of no CTC's raised by the choice $a_2=1/2$. If one argues that the dependence of $M_{\bar{r}}$ (and $E$) with the radial distance invalidates the argumentation on the string instability (based, for instance, in a similar argumentation of the one presented in the footnote 1), then it necessarily leads to the adoption of $a_2=1/2$ (with suitable choices for the other parameters) in order to avoid CTC's in the spacetime. This possibility is a genuine aspect of the analysis performed in framework analyzed. For instance, at least theoretically, it is intriguing that the energy (and $M_{\bar{r}}$ as well) does not depend on $\bar{r}$ in the Jordan frame for $\omega\rightarrow 0$.

We also verified that a similar spacetime to the one obtained can not describe the rotational curves of the analyzed galaxies in Ref. \cite{Rubin1980}, because, as $r$ increases, the observations indicate a tendency for a constant tangential velocity, but this is not attended.

We finalize stressing that the complete spacetime analysis, taking into account the interior solution, is under investigation and shall make explicit the hole performed by the BD parameter. To the best of our hope, a procedure akin to the one presented in Ref. \cite{Jensen1992A} working out the ballpoint pen model, along with appropriate junction conditions may lead to a solution free of naked singularities.

\begin{acknowledgments}
SMS thanks to CAPES for partial support. JMHS thanks to CNPq for partial support.
\end{acknowledgments}

\end{document}